\documentclass[a4paper]{article}
\RequirePackage[english]{babel}
\RequirePackage[latin1]{inputenc}
\RequirePackage[T1]{fontenc}
\RequirePackage{mathrsfs}
\RequirePackage{amsmath}
\RequirePackage{amssymb}
\RequirePackage{amsbsy}
\RequirePackage{bm}
\begin{document}
\title{\bf{Conformal Gravity with Electrodynamics\\ for Fermion Fields and their\\ Symmetry Breaking Mechanism}}
\author{Luca Fabbri\\
\footnotesize DIPTEM Sez. Metodi e Modelli Matematici, Universit\`{a} di Genova and \\
\footnotesize Dipartimento di Fisica, Universit{\`a} di Bologna \& INFN}
\date{}
\maketitle
\begin{abstract}
In this paper we consider an axial torsion to build metric-compatible connections in conformal gravity, with gauge potentials; the geometric background is filled with Dirac spinors: scalar fields with suitable potentials are added eventually. The system of field equations is worked out to have torsional effects converted into spinorial self-interactions: the massless spinors display self-interactions of a specific form that gives them the features they have in the non-conformal theory but with the additional character of renormalizability, and the mechanisms of generation of mass and cosmological constants become dynamical. As a final step we will address the cosmological constant and coincidence problems.
\end{abstract}
\begin{flushleft}
\textit{This paper is warmly dedicated to the memory of\\
Professor Mauro Francaviglia, founder of the\\
Italian Society of Gravitation SIGRAV.}
\end{flushleft}
\section*{Introduction}
When developing a geometry based on the most general spacetime coordinate invariance, both for its basic quantities and all of their derivatives, one is compelled to introduce an object called connection $\Gamma^{\alpha}_{\beta\mu}$ which in general is not symmetric in the two lower indices, so that $\Gamma^{\alpha}_{\mu\nu}\!-\!\Gamma^{\alpha}_{\nu\mu}\!=\!Q^{\alpha}_{\mu\nu}$ is different from zero and it turns out to be a tensor, called Cartan torsion tensor; despite that in general torsion is different from zero, there have been attempts to show that torsion should vanish if we want the principle of equivalence to realize an unambiguous geometrization of the gravitational field \cite{m-t-w}: even though it is true that for this to occur the vanishing of torsion is a sufficient condition, we need not require so much as the vanishing of the whole torsion since a completely antisymmetric torsion would be enough \cite{a-l,m-l,xy,so,F1}. Therefore in order for the principle of equivalence to unambiguously geometrize gravity a necessary and sufficient condition is the complete antisymmetry of Cartan torsion tensor.

Another aspect that a relativistic theory might display is the property of being conformally invariant: this property means that beside spacetime coordinate invariance also spacetime scale invariance is accounted; the feature of conformal symmetry is certainly one of the simplest and most intuitive among all possible symmetries that may characterize a given system; but scale symmetry in spacetime is also important because on the basis of such a constraint there is a single dynamical action that is selected out among all dynamical actions that would in principle be possible, as discussed by Weyl. With conformal invariance, the first point to settle is finding a way to implement conformal transformations for torsion, and as discussed in \cite{sh} this may be done either in terms of the \emph{strong} conformal transformation, for which torsion is assumed to have a general conformal transformation, or the \emph{weak} conformal transformation, for which torsion does not have a conformal transformation at all: the strong conformal transformation for torsion has been studied in \cite{fabbri}, where it has also been proven that such a transformation cannot be reduced to the weak conformal transformation for torsion, and so the case of weak conformal transformation for the torsion tensor must be studied independently; another reason is that strong conformal transformations are entirely loaded onto the trace vector part, and since in a conformal theory it is meaningless to require the vanishing of something that is not conformally invariant then it is impossible to have strong conformal transformations with a vanishing trace torsion vector, so that this case is incompatible with an irreducible torsion such as the one provided if torsion is completely antisymmetric, and therefore only the weak conformal transformations can be compatible with the completely antisymmetric torsion. Apart from this, another important reason to study weak conformal transformation for torsion is that conformal properties are purely metric concepts, thus independent on torsion: so it seems all too natural that a purely metric conformal transformation should leave the torsion tensor untouched. In the present paper, we therefore consider an axial torsion subject to no conformal transformation whatsoever.

It is important to remark a quite intriguing aspect: we have already mentioned that the complete antisymmetry of torsion makes it possible to have an unambiguous implementation of the principle of equivalence in order to geometrize gravity while conformal symmetry is related to the projective structure of the spacetime; an axial torsion tensor within a theory with conformal symmetry has the possibility to provide both the free-fall motion and the light-cone structures that are used in the Ehlers-Pirani-Schild construction \cite{e-p-s}.

The most important consequence of not neglecting torsion in a gravitational spacetime, eventually complemented with gauge fields, is that in so doing the underlying background is endowed with both torsion and curvature, beside all gauge strengths, so that one has all the elements that are needed, when such an underlying background is filled with matter fields with spin and energy, beside all currents, to couple torsion to spin much in the same way in which curvature is coupled to energy, like all gauge strengths are coupled to all currents, respectively, in the Poincar\'{e}-gauge theory \cite{h-h-k-n}: notice that a completely antisymmetric torsion is compatible with spin-$\frac{1}{2}$ spinors alone \cite{F2}. This is good since Dirac fields are the only fermionic matter we have ever observed.

The importance of scale symmetry in geometry lies on the fact that it has an analogous role to the one renormalizability has when one needs to fix the dynamics of the matter field content; this is intriguing because, as geometry is what encodes gravity, scale invariant geometries should be those that might ensure renormalizability of the gravitational dynamics; in \cite{s} Stelle proved this fact making conformal gravity a viable quantum gravity prototype. On the other hand the unambiguous geometrization of gravity via the principle of equivalence requires torsion to be completely antisymmetric, and as we have already discussed the axial torsion vector is by construction conformally invariant in an automatic way; conformal axial torsion gravity with gauge fields achieves the task of allowing the torsion-spin and curvature-energy, as well as the gauge strength-current, couplings, with torsion and gravity, beside gauge fields, being renormalizable as desired, and additionally this is accomplished for the case of spin-$\frac{1}{2}$ massless spinor fields. As discussed in \cite{Fabbri:2011ha} in general conformal geometries Dirac fermions might have ill-defined structure, but we will see here that in the present conformal geometry Dirac fermions emerge consistently.

However, our discussion will not be focused on the Dirac fermions alone, and as a final ingredient, we will consider the real scalar field, not only as a matter of completeness, but also because the scalar field possesses a potential providing a symmetry breaking mechanism that, in the conformal gravitation, is absolutely necessary: in fact a theory of gravity with conformal invariance suffers the problem that it seems to be incompatible with a universe that does not appear to possess such a symmetry; this problem is solved when a mechanism of conformal symmetry breaking is given. The scalar field, with its peculiar conformal behaviour, possesses all that is needed to have the conformal symmetry broken by means of gravitational actions in a dynamical way; furthermore, we will see that the scalar field will contribute to the spin density and therefore to the torsion, and this surprising fact will render the conformal symmetry breaking due gravity with torsion even more dynamical. This dynamical conformal symmetry breaking will have unexpected consequences for the fields involved.

In this paper, we discuss these consequences focusing on applications to the standard models of particle physics and cosmology, and in particular to the generation of the Higgs mass and the cosmological constant, addressing both cosmological constant and coincidence problem at once.
\section{Kinematic Background}
In the present construction we refer for the introduction of the general formalism and definitions to \cite{Fabbri:2011ha}, and here we will review the most important concepts.

We recall that the completely antisymmetric torsion is an axial torsion since once the completely antisymmetric Levi-Civita density $\varepsilon_{\alpha\nu\pi\sigma}$ is introduced then one may write the completely antisymmetric torsion $Q_{\rho\beta\mu}$ in terms of an axial torsion $W^{\theta}$ according to $Q_{\alpha\mu\nu}\!=\!\varepsilon_{\alpha\mu\nu\sigma}W^{\sigma}$ and thus, taking the Levi-Civita connection $\Lambda^{\alpha}_{\beta\mu}$ we can write $\Gamma^{\alpha}_{\beta\mu}
\!=\!\frac{1}{2}g^{\alpha\rho}\varepsilon_{\rho\beta\mu\sigma}W^{\sigma}\!+\!\Lambda^{\alpha}_{\beta\mu}$ as the most general decomposition of the connection we will consider in this paper, and as we have anticipated, the torsion $Q^{\alpha}_{\phantom{\alpha}\beta\mu}$ originally defined with the first upper index and the last two lower indices has no conformal transformation, while the metric has the usual form of conformal transformation; the most general connection defines the most general covariant derivative $D_{\mu}$ while the Levi-Civita connection defines the Levi-Civita covariant derivative $\nabla_{\mu}$ for which we have that the following relationships $D_{\mu}g_{\alpha\rho}\!=\!\nabla_{\mu}g_{\alpha\rho}\!=\!0$ and $D_{\mu}\varepsilon_{\rho\beta\nu\sigma}\!=\!\nabla_{\mu}\varepsilon_{\rho\beta\nu\sigma}\!=\!0$ hold and they are called metric-compatibility conditions: these conditions and the complete antisymmetry of torsion imply the connection to have a single symmetric part entirely written in terms of the metric, so that the gravitational information, which according to Weyl theorem and the principle of equivalence had to be in the symmetric part of the connection, is forced into the metric, and so the interpretation of the gravitational field as a geometric entity is compulsory, and one does not need to employ any additional prescription in order to declare where the gravitational information is actually stored, as shown in \cite{a-l,m-l,xy,so,F1}.

Now an equivalent formalism may be introduced, in which the metric is written as $g_{\alpha\nu}=e_{\alpha}^{p}e_{\nu}^{i} \eta_{pi}$ in terms of the Minkowskian metric $\eta_{ij}$ and a basis of vierbein $e_{\alpha}^{i}$ and such that with them, the connection can be transformed into the spin-connection $\omega^{i}_{\phantom{i}p\alpha}\!=\!
e^{i}_{\sigma}(\Gamma^{\sigma}_{\rho\alpha}e^{\rho}_{p}\!+\partial_{\alpha}e^{\sigma}_{p})$ such that $\omega^{ip}_{\phantom{ip}\alpha}\!=\!-\omega^{pi}_{\phantom{pi}\alpha}$ stating the antisymmetry of the spin-connection; these last two relationships are related to the conditions $D_{\mu}e_{\alpha}^{k}\!=\!\nabla_{\mu}e_{\alpha}^{k}\!=\!0$ and $D_{\mu}\eta_{ij}\!=\!\nabla_{\mu}\eta_{ij}\!=\!0$ known as coordinate-Lorentz compatibility conditions, since they establish that the previously defined coordinate formalism with Greek indices and the present formalism with Latin indices are such that at any differential level they are equivalent at all.

The reason for which such an equivalent but different formalism has been introduced is that with it the most general coordinate transformations are replaced by the Lorentz transformations, whose explicit form can be rewritten in terms of other representations such as the complex one; since complex representations of the Lorentz transformation act on complex fields, the geometry of complex fields is given in terms of a gauge connection $A_{\mu}$ that will serve to introduce the gauge covariant derivatives, as usual in abelian gauge theories.

Now, an explicit representation of the complex Lorentz transformation can be achieved through the introduction of the $\boldsymbol{\gamma}_{a}$ matrices verifying the Clifford algebra $\{\boldsymbol{\gamma}_{a},\boldsymbol{\gamma}_{b}\}\!=\!2\boldsymbol{\mathbb{I}}\eta_{ab}$ from which we define $\boldsymbol{\sigma}_{ab}\!=\!\frac{1}{4}[\boldsymbol{\gamma}_{a},\boldsymbol{\gamma}_{b}]$ such that they verify the conditions $\{\boldsymbol{\gamma}_{a},\boldsymbol{\sigma}_{bc}\}\!=\!i\varepsilon_{abcd} \boldsymbol{\gamma}\boldsymbol{\gamma}^{d}$ as the complex generators of the complex representation of the Lorentz algebra, and from which we define the most general spinorial-connection as $\boldsymbol{\Omega}_{\rho}
\!=\!\frac{1}{2}\omega^{ij}_{\phantom{ij}\rho}\boldsymbol{\sigma}_{ij}\!+\!iqA_{\rho}\mathbb{I}$ in terms of the parameter $q$ known as the charge of the spin-$\frac{1}{2}$ spinorial field, and as it is expected this spinorial connection can now be used to define the spinorial covariant derivatives given in the notation $\boldsymbol{D}_{\mu}$ as usual; then we have $\boldsymbol{D}_{\mu}\boldsymbol{\gamma}_{a}\!=\!\boldsymbol{\nabla}_{\mu}\boldsymbol{\gamma}_{a} \!=\!0$ automatically.

We notice that the introduction of vierbein and spin-connection is what ultimately allows the definition of the spinor fields, and so these structures are essential for the definition of spin structures, as explained in \cite{Fatibene:1998aa,Fatibene:1996mc}.

From the connection and its torsionless Levi-Civita connection we define
\begin{eqnarray}
&G^{\rho}_{\phantom{\rho}\xi\mu\nu}
=\partial_{\mu}\Gamma^{\rho}_{\xi\nu}-\partial_{\nu}\Gamma^{\rho}_{\xi\mu}
+\Gamma^{\rho}_{\sigma\mu}\Gamma^{\sigma}_{\xi\nu}
-\Gamma^{\rho}_{\sigma\nu}\Gamma^{\sigma}_{\xi\mu}
\label{curvature}\\
&R^{\rho}_{\phantom{\rho}\xi\mu\nu}
=\partial_{\mu}\Lambda^{\rho}_{\xi\nu}-\partial_{\nu}\Lambda^{\rho}_{\xi\mu}
+\Lambda^{\rho}_{\sigma\mu}\Lambda^{\sigma}_{\xi\nu}
-\Lambda^{\rho}_{\sigma\nu}\Lambda^{\sigma}_{\xi\mu}
\label{metriccurvature}
\end{eqnarray}
with decomposition given by
\begin{eqnarray}
\nonumber
&G_{\rho\xi\mu\nu}=R_{\rho\xi\mu\nu}
+\frac{1}{2}\left(\varepsilon_{\rho\xi\nu\alpha}\nabla_{\mu}W^{\alpha}
-\varepsilon_{\rho\xi\mu\alpha}\nabla_{\nu}W^{\alpha}\right)-\\
&-\frac{1}{4}(W_{\rho}W_{[\mu}g_{\nu]\xi}-W_{\xi}W_{[\mu}g_{\nu]\rho})
+\frac{1}{8}W^{2}(g_{\rho[\mu}g_{\nu]\xi}-g_{\xi[\mu}g_{\nu]\rho})
\label{decomposition}
\end{eqnarray}
in terms of the axial torsion, known as Riemann and purely metric Riemann curvature tensors; they are antisymmetric in both the first and second pair of indices, and so with contraction $G^{\rho}_{\phantom{\rho}\mu\rho\nu}\!=\!G_{\mu\nu}$ and $R^{\rho}_{\phantom{\rho}\mu\rho\nu}\!=\!R_{\mu\nu}$ themselves with contractions given by $G_{\eta\nu}g^{\eta\nu}\!=\!G$ and $R_{\eta\nu}g^{\eta\nu}\!=\!R$ as it is usual: curvatures with the same symmetries for indices transposition, but irreducible, are given by
\begin{eqnarray}
&W_{\alpha\beta\mu\nu}=G_{\alpha\beta\mu\nu}
-\frac{1}{2}(G_{\alpha[\mu}g_{\nu]\beta}-G_{\beta[\mu}g_{\nu]\alpha})
+\frac{1}{12}G(g_{\alpha[\mu}g_{\nu]\beta}-g_{\beta[\mu}g_{\nu]\alpha})
\label{irreduciblecurvature}\\
&C_{\alpha\beta\mu\nu}=R_{\alpha\beta\mu\nu}
-\frac{1}{2}(R_{\alpha[\mu}g_{\nu]\beta}-R_{\beta[\mu}g_{\nu]\alpha})
+\frac{1}{12}R(g_{\alpha[\mu}g_{\nu]\beta}-g_{\beta[\mu}g_{\nu]\alpha})
\label{irreduciblemetriccurvature}
\end{eqnarray}
related through the decomposition
\begin{eqnarray}
&W_{\alpha\beta\mu\nu}=C_{\alpha\beta\mu\nu}
+\frac{1}{2}\left[\varepsilon_{\alpha\beta\theta[\mu}g_{\nu]\rho}
-\frac{1}{2}\left(\varepsilon_{\alpha\rho\theta[\mu}g_{\nu]\beta}
-\varepsilon_{\beta\rho\theta[\mu}g_{\nu]\alpha}\right)\right]\nabla^{\rho}W^{\theta}
\label{irreducibledecomposition}
\end{eqnarray}
in which due to the derivatives of the axial torsion, $W_{\alpha\beta\mu\nu}$ is not conformally invariant even though $C_{\alpha\beta\mu\nu}$ is conformally invariant, and this is what is known as the Weyl curvature conformal tensor. The reason for this is seen in the implicit presence of torsion within the curvature, and in particular to the fact that there are derivatives of torsion, beside squared torsion terms, in the curvature, and thus in the irreducible curvature itself: when conformal transformations are used, the derivatives of torsion always transform, spoiling for $W_{\alpha\beta\mu\nu}$ the conformal invariance that instead for $C_{\alpha\beta\mu\nu}$ was ensured; thus a solution can be obtained by adding to the curvature $G_{\alpha\beta\mu\nu}$ some terms with torsion getting a modified torsional-curvature $M_{\alpha\beta\mu\nu}$ in which the simultaneous presence of torsion implicitly through $G_{\alpha\beta\mu\nu}$ and explicitly added would provide the cancellation of all extra terms during the conformal transformation, yielding a tensor $M_{\alpha\beta\mu\nu}$ whose irreducible part $T_{\alpha\beta\mu\nu}$ is conformal. In \cite{fabbri} the solution we found for strong conformal transformations of torsion was adding to the tensor $G_{\alpha\beta\mu\nu}$ \emph{squared} torsional terms, to get $M_{\alpha\beta\mu\nu}$ in which the non-conformal transformation of the derivatives of torsion and that of the squared torsion contributions cancelled away, so to have $T_{\alpha\beta\mu\nu}$ conformally invariant; but in the present paper the axial torsion has weak conformal transformation, that is no conformal transformation, which cannot provide the extra pieces needed for such a cancellation to occur, and therefore to $G_{\alpha\beta\mu\nu}$ we have to add beside \emph{squared} terms also \emph{derivatives} of torsion, and when this is done in the most general way in terms of the $a$, $b$ and $c$ parameters, we have that the modified tensor
\begin{eqnarray}
\nonumber
&M_{\alpha\beta\mu\nu}=G_{\alpha\beta\mu\nu}
+\frac{c}{2}(\varepsilon_{\alpha\beta\mu\theta}D_{\nu}W^{\theta}
-\varepsilon_{\alpha\beta\nu\theta}D_{\mu}W^{\theta})+\\
&+\frac{a}{4}(W_{\alpha}W_{[\mu}g_{\nu]\beta}-W_{\beta}W_{[\mu}g_{\nu]\alpha})
+\frac{b}{8}W^{2}(g_{\alpha[\mu}g_{\nu]\beta}-g_{\beta[\mu}g_{\nu]\alpha})
\end{eqnarray}
has irreducible part given by
\begin{eqnarray}
\nonumber
&T_{\alpha\beta\mu\nu}=C_{\alpha\beta\mu\nu}
\end{eqnarray}
and thus conformally invariant, for any value of the $a$ and $b$ parameters and with the condition $c\!=\!1$ that must be taken, showing that no squared term can be useful and that there is a single choice for which all the derivatives of the axial torsion cancel away. However in this way, although $T_{\alpha\beta\mu\nu}$ is conformally invariant nevertheless the fact that it reduces to $C_{\alpha\beta\mu\nu}$ implies that the correction is somewhat trivial; that is in the case of axial torsion with no conformal transformations any correction to the Riemann curvature is such that its irreducible part is either not conformally invariant or it reduces to the Weyl tensor and so it is conformally invariant in a trivial way. Or equivalently, the Weyl tensor is the most general irreducible tensor that is also conformally invariant.

Before proceeding, we have also to recall that for the gauge connection the gauge curvature is defined as $F_{\mu\nu}\!=\!\nabla_{\mu}A_{\nu}\!-\!\nabla_{\nu}A_{\mu}$ antisymmetric and thus irreducible, and as the gauge connection is usually assumed to have a trivial conformal transformation it follows that the gauge curvature is conformal.

What transforms according to the complex representation of the Lorentz transformation, that is the spinorial transformation, is defined to be the spinor field $\psi$ whose conformal transformation is given by $\psi\rightarrow\sigma^{-\frac{3}{2}}\psi$ as usual.

Finally, the real scalar field $\phi$ has scaling $\sigma^{-1}$ as usual: notice that according to the usual way to count degrees of freedom, massless real scalars should have no degree of freedom at all; in conformal models this is well understood since according to their conformal transformations they may always be conformally transformed away. However, we will see that this does not amount to have no scalar field at all, and in fact conformally transforming the massless real scalar away eventually provokes the dynamical symmetry breaking.
\section{Dynamical Action}
We have seen in the previous section that efforts to follow the line of \cite{fabbri,Fabbri:2011ha} and get a curvature that is conformally invariant in the case of axial torsion with no conformal transformations bring the most general of such curvatures back to the torsionless curvature with conformal invariance known as Weyl tensor, and so Cartan torsion and Weyl curvature are independently conformally invariant. 

However, if in \cite{Fabbri:2011ha} the strong conformal transformations allowed us to employ torsion to modify the Weyl curvature as to achieve its conformal invariance but also forbade torsion to explicitly appear in the action without spoiling conformal invariance, here the weak conformal transformation that keeps Cartan torsion and Weyl curvature separated allows Cartan torsion to explicitly enter beside Weyl curvature in actions maintaining conformal invariance; it is easy to see that there is no possible product of Cartan torsions and Weyl curvatures that could enter in the action preserving its conformal invariance, and Cartan torsion and Weyl curvature will be separated also dynamically. To define the dynamical features of the model, we have to take into account the fact that no derivative of the axial torsion should enter the action: the first reason for this is that we have already seen that derivatives of the axial torsion are not in general conformally invariant, although this problem might be circumvented by the fact that there are many of such terms, each entering with its own coupling factor, for which a fine-tuning may restore conformal invariance; a second reason for this consists in the fact that any derivative of torsion would result into dynamical torsional field equations, allowing torsion to propagate out of matter, but there are stringent limits for the presence of torsion in vacuum. If we want no derivative of torsion in the action, then $\left|W^{\alpha}W_{\alpha}\right|^{2}$ is the only axial torsion term that defines an action with conformal invariance; the term $C_{\alpha\beta\mu\nu}C^{\alpha\beta\mu\nu}$ defines the action with conformal invariance for gravitation as usual: therefore we have that
\begin{eqnarray}
\nonumber
&S_{\mathrm{gravity}}=\int\frac{1}{4}(3kW^{\alpha}W_{\alpha}W^{\rho}W_{\rho}
+C_{\alpha\beta\mu\nu}C^{\alpha\beta\mu\nu})\sqrt{|g|}dV
\label{actiongravitation}
\end{eqnarray}
with torsion constant $k$ and with gravitational constant normalized to unity, and under the hypotheses above this is the most general conformal action.

For the gauge field the term $F_{\mu\nu}F^{\mu\nu}$ is the only one to be included as
\begin{eqnarray}
&S_{\mathrm{electrodynamics}}
=\int\left(-\frac{1}{4}F_{\alpha\beta}F^{\alpha\beta}\right)\sqrt{|g|}dV
\label{actionelectrodynamics}
\end{eqnarray}
and this is the electrodynamic action, and it is conformally invariant.

The Dirac spinorial action is given as usual by
\begin{eqnarray}
&S_{\mathrm{matter}}=\int[\frac{i}{2}(\overline{\psi}\boldsymbol{\gamma}^{\rho}\boldsymbol{D}_{\rho}\psi
-\boldsymbol{D}_{\rho}\overline{\psi}\boldsymbol{\gamma}^{\rho}\psi)]|e|dV
\label{actionmatter}
\end{eqnarray}
and this is the material action that in the massless case is conformally invariant.

For the real scalar action we have to notice an important fact: due to the second-order derivative of the scalar, this field is not naturally conformally invariant; therefore some additional terms must be added to the dynamical term in order to obtain the conformal invariance of the whole action: such extra terms consist first of all of the $R\phi^{2}$ term as it is known; however in presence of the axial torsion also the $W^{2}\phi^{2}$ term is present: the entire scalar action is
\begin{eqnarray}
&S_{\mathrm{scalar}}=\int(\nabla_{\rho}\phi\nabla^{\rho}\phi\!+\!\frac{1}{6}R\phi^{2} \!-\!pW_{\alpha}W^{\alpha}\phi^{2}\!-\!\frac{\lambda}{8}\phi^{4})|e|dV
\label{actionSB}
\end{eqnarray}
where the quartic potential has also been added, and it is given in terms of the two coupling constants $p$ and $\lambda$ characterizing this conformally invariant action.

Once all is summed up, the total action is given by
\begin{eqnarray}
\nonumber
&S=\int[\frac{3}{4}kW^{\alpha}W_{\alpha}W^{\mu}W_{\mu}
+\frac{1}{4}C_{\alpha\beta\mu\nu}C^{\alpha\beta\mu\nu}-\frac{1}{4}F_{\alpha\beta}F^{\alpha\beta}+\\
\nonumber
&+\frac{i}{2}(\overline{\psi}\boldsymbol{\gamma}^{\rho}\boldsymbol{\nabla}_{\rho}\psi
-\boldsymbol{\nabla}_{\rho}\overline{\psi}\boldsymbol{\gamma}^{\rho}\psi)
-\frac{3}{4}\overline{\psi}\boldsymbol{\gamma}_{\rho}\boldsymbol{\gamma}\psi W^{\rho}+\\
&+\nabla_{\rho}\phi\nabla^{\rho}\phi+\frac{1}{6}R\phi^{2}-pW_{\alpha}W^{\alpha}\phi^{2}
-\frac{\lambda}{8}\phi^{4}-Y\overline{\psi}\psi\phi]|e|dV
\label{action}
\end{eqnarray}
where the additional Yukawa term with constant $Y$ has been added, and up to the coupling constants this conformally invariant action is defined uniquely.

By varying with respect to the fields involved one gets the axial torsion-spin and irreducible curvature-energy field equations given by the expressions
\begin{eqnarray}
&kW^{2}W_{\rho}=\frac{1}{4}\overline{\psi}\boldsymbol{\gamma}_{\rho}\boldsymbol{\gamma}\psi
+\frac{2p}{3}\phi^{2}W_{\rho}
\label{torsionequations}
\end{eqnarray}
and
\begin{eqnarray}
\nonumber
&-kW^{2}(\frac{1}{4}g^{\mu\alpha}W^{2}-W^{\mu}W^{\alpha})+(\frac{1}{4}g^{\mu\alpha}C^{2}
-C^{\theta\sigma\rho\mu}C_{\theta\sigma\rho}^{\phantom{\theta\sigma\rho}\alpha})-\\
\nonumber
&-(C^{\mu\beta\alpha\nu}R_{\beta\nu}+2\nabla_{\beta}\nabla_{\nu}C^{\mu\beta\alpha\nu})
-(\frac{1}{4}g^{\mu\alpha}F^{2}-F^{\mu\rho}F^{\alpha}_{\phantom{\alpha}\rho})=\\
\nonumber
&=\frac{i}{4}(\overline{\psi}\boldsymbol{\gamma}^{\mu}\boldsymbol{\nabla}^{\alpha}\psi
-\boldsymbol{\nabla}^{\alpha}\overline{\psi}\boldsymbol{\gamma}^{\mu}\psi
+\overline{\psi}\boldsymbol{\gamma}^{\alpha}\boldsymbol{\nabla}^{\mu}\psi
-\boldsymbol{\nabla}^{\mu}\overline{\psi}\boldsymbol{\gamma}^{\alpha}\psi)+\\
\nonumber
&+\frac{1}{4}
(\frac{1}{2}\overline{\psi}\boldsymbol{\gamma}^{\alpha}\boldsymbol{\gamma}\psi W^{\mu}+
\frac{1}{2}\overline{\psi}\boldsymbol{\gamma}^{\mu}\boldsymbol{\gamma}\psi W^{\alpha}
-g^{\alpha\mu}\overline{\psi}\boldsymbol{\gamma}^{\rho}\boldsymbol{\gamma}\psi W_{\rho})+\\
\nonumber
&+2(\nabla^{\alpha}\phi\nabla^{\mu}\phi
-\frac{1}{2}g^{\alpha\mu}\nabla_{\rho}\phi\nabla^{\rho}\phi)
+\frac{1}{3}(g^{\alpha\mu}\nabla^{2}\phi^{2}-\nabla^{\alpha}\nabla^{\mu}\phi^{2})+\\
&+\frac{1}{3}(R^{\alpha\mu}-\frac{1}{2}g^{\alpha\mu}R)\phi^{2}
+\frac{2p}{3}(W^{\alpha}W^{\mu}+\frac{1}{2}g^{\alpha\mu}W^{2})\phi^{2}
+g^{\alpha\mu}\frac{\lambda}{8}\phi^{4}
\end{eqnarray}
with the gauge strength-current field equations
\begin{eqnarray}
&\nabla_{\rho}F^{\rho\mu}=q\overline{\psi}\boldsymbol{\gamma}^{\mu}\psi
\end{eqnarray}
for the geometric sector; the fermionic field equations are
\begin{eqnarray}
&i\boldsymbol{\gamma}^{\mu}\boldsymbol{\nabla}_{\mu}\psi
-\frac{3}{4}W_{\sigma}\boldsymbol{\gamma}^{\sigma}\boldsymbol{\gamma}\psi-Y\phi\psi=0
\end{eqnarray}
for the fermionic sector. The scalar field equations are
\begin{eqnarray}
&\nabla^{2}\phi-\frac{1}{6}R\phi+pW^{2}\phi
+\frac{\lambda}{4}\phi^{2}\phi+\frac{Y}{2}\overline{\psi}\psi=0
\end{eqnarray}
for which the vacuum giving stable stationary breakdown of the symmetry is
\begin{eqnarray}
&\left.\left(\frac{\lambda}{4}\phi^{2}+pW^{2}-\frac{1}{6}R\right)\right|_{\mathrm{v}}=0
\end{eqnarray}
as it is clear after performing a straightforward calculation.

As a final remark we have to notice that in the previously elaborated conformal model with strong transformation for torsion the field equations coupled a mixture of curvature and torsion to the spin and to the energy, so that in particular there was no completely antisymmetric torsion matching the completely antisymmetric spin of the Dirac field \cite{Fabbri:2011ha}, whereas in this conformal gravity with no transformation for torsion there is no such mixing of degrees of freedom, and consequently the field equations couple torsion to the spin and curvature to the energy, and in particular the axial torsion and the axial spin of the Dirac field correspond perfectly: then in all field equations the number of degrees of freedom and independent fields matches consistently; in particular we have to notice that both the spinorial and the scalar field equations have characteristic equations given by the simplest $n^{2}\!=\!0$ showing that the characteristic surfaces propagate inside the light-cone respecting 
causality \cite{F2}.
\subsection{The Non-Linear Potential}
In this subsection we will perform the decomposition of the field equations, but in order to proceed it is essential that the field equations for the torsion-spin coupling be inverted in such a way that torsion is given in terms of the spin density of the matter fields: to do so consider field equations (\ref{torsionequations}) squared as
\begin{eqnarray}
&(kW^{2})^{3}-\frac{4p}{3}\phi^{2}(kW^{2})^{2}+\frac{4p^{2}}{9}\phi^{4}(kW^{2})
+\frac{k}{16}\overline{\psi}\boldsymbol{\gamma}_{i}\psi\overline{\psi}\boldsymbol{\gamma}^{i}\psi=0
\end{eqnarray}
which can always be solved via Cardano cubic formula, and thus by putting
\begin{eqnarray}
&\Delta\equiv4\frac{729k\overline{\psi}\boldsymbol{\gamma}_{i}\psi \overline{\psi}\boldsymbol{\gamma}^{i}\psi}{512p^{3}\phi^{6}}(1+\frac{729k\overline{\psi}\boldsymbol{\gamma}_{i}\psi 
\overline{\psi}\boldsymbol{\gamma}^{i}\psi}{512p^{3}\phi^{6}})
\label{discriminant}
\end{eqnarray}
known as discriminant we have that solutions are given by
\begin{eqnarray}
&\frac{9kW^{2}}{2p\phi^{2}}=
2\!-\!u[(1\!+\!\Delta)^{\frac{1}{2}}\!+\!\Delta^{\frac{1}{2}}]^{\frac{1}{3}}
\!-\!u^{\ast}[(1\!+\!\Delta)^{\frac{1}{2}}\!-\!\Delta^{\frac{1}{2}}]^{\frac{1}{3}}
\end{eqnarray}
where $u$ are the three complex cubic roots of the unity, therefore corresponding to the three complex solutions in general: having these, the field equations for the torsion-spin coupling given above can finally be inverted as to get
\begin{eqnarray}
&W_{\rho}=-\frac{9}{8p\phi^{2}\left[1\!+\!u[(1\!+\!\Delta)^{\frac{1}{2}}\!
+\!\Delta^{\frac{1}{2}}]^{\frac{1}{3}}\!+\!u^{\ast}[(1\!+\!\Delta)^{\frac{1}{2}}\!
-\!\Delta^{\frac{1}{2}}]^{\frac{1}{3}}\right]}
\overline{\psi}\boldsymbol{\gamma}_{\rho}\boldsymbol{\gamma}\psi
\end{eqnarray}
used to substitute torsion with the spin of matter fields. When this is done also the curvature-energy coupling field equations can be inverted resulting in
\begin{eqnarray}
\nonumber
&(\frac{1}{4}g_{\mu\alpha}C^{2}
-C^{\theta\sigma\rho}_{\phantom{\theta\sigma\rho}\mu}C_{\theta\sigma\rho\alpha}-C_{\mu\beta\alpha\nu}R^{\beta\nu}-2\nabla^{\beta}\nabla^{\nu}C_{\mu\beta\alpha\nu})-\\
\nonumber
&-(\frac{1}{4}g_{\mu\alpha}F^{2}-F_{\mu\rho}F_{\alpha}^{\phantom{\alpha}\rho})
=\frac{i}{4}\left(\overline{\psi}\boldsymbol{\gamma}_{(\mu}\boldsymbol{\nabla}_{\alpha)}\psi
-\boldsymbol{\nabla}_{(\alpha}\overline{\psi}\boldsymbol{\gamma}_{\mu)}\psi\right)+\\
\nonumber
&+2(\nabla_{\alpha}\phi\nabla_{\mu}\phi
-\frac{1}{2}g_{\alpha\mu}\nabla_{\rho}\phi\nabla^{\rho}\phi)
+\frac{1}{3}(g_{\alpha\mu}\nabla^{2}\phi^{2}-\nabla_{\alpha}\nabla_{\mu}\phi^{2})+\\
\nonumber
&+\frac{1}{3}(R_{\alpha\mu}-\frac{1}{2}g_{\alpha\mu}R)\phi^{2}+\\
\nonumber
&+g_{\alpha\mu}[\frac{\lambda}{8}\!+\!\frac{p^{2}}{27k}
[4\!+\!u[(1\!+\!\Delta)^{\frac{1}{2}}\!+\!\Delta^{\frac{1}{2}}]^{\frac{1}{3}}
\!+\!u^{\ast}[(1\!+\!\Delta)^{\frac{1}{2}}\!-\!\Delta^{\frac{1}{2}}]^{\frac{1}{3}}]\cdot\\
&\cdot[2\!-\!u[(1\!+\!\Delta)^{\frac{1}{2}}\!+\!\Delta^{\frac{1}{2}}]^{\frac{1}{3}}
\!-\!u^{\ast}[(1\!+\!\Delta)^{\frac{1}{2}}\!-\!\Delta^{\frac{1}{2}}]^{\frac{1}{3}}]]\phi^{4}
\end{eqnarray}
with non-linear contributions while the gauge strength-current field equations
\begin{eqnarray}
&\nabla_{\rho}F^{\rho\mu}=q\overline{\psi}\boldsymbol{\gamma}^{\mu}\psi
\end{eqnarray}
are unmodified by the non-linearities instead; the fermionic field equations are
\begin{eqnarray}
&i\boldsymbol{\gamma}^{\mu}\boldsymbol{\nabla}_{\mu}\psi
-\frac{27}{32p\phi^{2}\left[1\!+\!u[(1\!+\!\Delta)^{\frac{1}{2}}\!
+\!\Delta^{\frac{1}{2}}]^{\frac{1}{3}}\!+\!u^{\ast}[(1\!+\!\Delta)^{\frac{1}{2}}\!
-\!\Delta^{\frac{1}{2}}]^{\frac{1}{3}}\right]}\overline{\psi}\boldsymbol{\gamma}_{\rho}\psi \boldsymbol{\gamma}^{\rho}\psi\!-\!Y\phi\psi\!=\!0
\end{eqnarray}
with non-linear interactions of the fermion fields. The scalar field equations are
\begin{eqnarray}
&\nabla^{2}\phi\!-\![\frac{1}{6}R\!+\!\frac{81\overline{\psi}\boldsymbol{\gamma}_{\rho}\psi
\overline{\psi}\boldsymbol{\gamma}^{\rho}\psi}{64p\phi^{4}
\left[1\!+\!u[(1\!+\!\Delta)^{\frac{1}{2}}\!+\!\Delta^{\frac{1}{2}}]^{\frac{1}{3}}\!
+\!u^{\ast}[(1\!+\!\Delta)^{\frac{1}{2}}\!-\!\Delta^{\frac{1}{2}}]^{\frac{1}{3}}\right]^{2}}
\!-\!\frac{\lambda}{4}\phi^{2}]\phi\!+\!\frac{Y}{2}\overline{\psi}\psi\!=\!0
\end{eqnarray}
also with non-linear interactions of the scalar field for which the vacuum that produces a stable stationary breakdown of the symmetry is given by
\begin{eqnarray}
&\left.\left[\frac{\lambda}{2}\phi^{2}-\frac{81\overline{\psi}\boldsymbol{\gamma}_{\rho}\psi
\overline{\psi}\boldsymbol{\gamma}^{\rho}\psi}{32p\phi^{4}\left[1\!
+\!u[(1\!+\!\Delta)^{\frac{1}{2}}\!+\!\Delta^{\frac{1}{2}}]^{\frac{1}{3}}\!
+\!u^{\ast}[(1\!+\!\Delta)^{\frac{1}{2}}\!-\!\Delta^{\frac{1}{2}}]^{\frac{1}{3}}\right]^{2}}-\frac{1}{3}R\right]\right|_{\mathrm{v}}=0
\end{eqnarray}
showing that the breaking of the symmetry through a dynamical mechanisms is possible not only in the case mediated by the gravitational field but also for a Ricci flat spacetime due to the torsionally-induced self-interactions of matter.

Let us now discuss the possibilities we have: among the three complex solutions given in general one is however always real and if $kp\!>\!0$ or in the case in which $kp\!<\!0$ but $|512p^{3}\phi^{6}|\!<\!|729k\overline{\psi}\boldsymbol{\gamma}_{\mu}\psi \overline{\psi}\boldsymbol{\gamma}^{\mu}\psi|$ then the discriminant is positive and the real root is unique, and we have only one solution; then in the case where $kp\!<\!0$ and $|512p^{3}\phi^{6}|\!\geqslant\!|729k\overline{\psi}\boldsymbol{\gamma}_{\mu}\psi \overline{\psi}\boldsymbol{\gamma}^{\mu}\psi|$ the discriminant is non-positive and all three roots are real, so that we have all three solutions, although in the case of equality two of them coincide, and so there are only two independent solutions, while in the case of strict inequality each of them differs from any other, and thus all three are distinct independent solutions. But nevertheless, these last situations are special, because for them it would always be possible for an assigned value of the scalar to have fermions with a density that is low enough for the solutions to become complex; to avoid such possibility, whatever matter fields are, we have to require $kp\!>\!0$ to hold as a general constraint, and thus this is what we shall do. Notice that having these two constants with the same sign still leaves us with the two alternatives about which  one of the two signs we choose, and therefore we are still left with two alternative models.

From now on we neglect the gravitational and electrodynamic field so to focus on the fermionic and scalar matter fields alone, and we take into account the condition $kp\!>\!0$ for which the discriminant (\ref{discriminant}) is always positive and the root that is always real comes for the choice $u\!=\!1$ as known: when this is done we get that the fermionic field equations are written according to the form
\begin{eqnarray}
&i\boldsymbol{\gamma}^{\mu}\boldsymbol{\nabla}_{\mu}\psi
-\frac{27}{32p\phi^{2}\left[1\!+\![(1\!+\!\Delta)^{\frac{1}{2}}\!
+\!\Delta^{\frac{1}{2}}]^{\frac{1}{3}}\!+\![(1\!+\!\Delta)^{\frac{1}{2}}\!
-\!\Delta^{\frac{1}{2}}]^{\frac{1}{3}}\right]}\overline{\psi}\boldsymbol{\gamma}_{\rho}\psi \boldsymbol{\gamma}^{\rho}\psi\!-\!Y\phi\psi\!=\!0
\end{eqnarray}
while the scalar field equations are given by
\begin{eqnarray}
&\nabla^{2}\phi\!-\!\left[\frac{81\overline{\psi}\boldsymbol{\gamma}_{\rho}\psi
\overline{\psi}\boldsymbol{\gamma}^{\rho}\psi}{64p\phi^{4}
\left[1\!+\![(1\!+\!\Delta)^{\frac{1}{2}}\!+\!\Delta^{\frac{1}{2}}]^{\frac{1}{3}}\!
+\![(1\!+\!\Delta)^{\frac{1}{2}}\!-\!\Delta^{\frac{1}{2}}]^{\frac{1}{3}}\right]^{2}}
\!-\!\frac{\lambda}{4}\phi^{2}\right]\phi\!+\!\frac{Y}{2}\overline{\psi}\psi\!=\!0
\end{eqnarray}
with stable stationary breakdown of the symmetry given when
\begin{eqnarray}
&\left.\left[\lambda\phi^{2}-\frac{81\overline{\psi}\boldsymbol{\gamma}_{\rho}\psi
\overline{\psi}\boldsymbol{\gamma}^{\rho}\psi}{16p\phi^{4}\left[1\!+\![(1\!+\!\Delta)^{\frac{1}{2}}
\!+\!\Delta^{\frac{1}{2}}]^{\frac{1}{3}}\!+\![(1\!+\!\Delta)^{\frac{1}{2}}
\!-\!\Delta^{\frac{1}{2}}]^{\frac{1}{3}}\right]^{2}}\right]\right|_{\mathrm{v}}=0
\end{eqnarray}
where we will set $\phi^{2}|_{\mathrm{v}}=v^{2}$ as the scalar vacuum constant value, and we will proceed in studying the two limits given by the approximation of high-density fermion fields compared to the scalar and the complementary approximation of low-density fermion fields compared to the scalar: in the former case we have
\begin{eqnarray}
&i\boldsymbol{\gamma}^{\mu}\boldsymbol{\nabla}_{\mu}\psi
-\left(\frac{27}{256k\overline{\psi}\boldsymbol{\gamma}_{i}\psi \overline{\psi}\boldsymbol{\gamma}^{i}\psi}\right)^{\frac{1}{3}}
\overline{\psi}\boldsymbol{\gamma}_{\rho}\psi \boldsymbol{\gamma}^{\rho}\psi-Yv\psi\approx0
\end{eqnarray}
while the scalar field equation is given by
\begin{eqnarray}
&\nabla^{2}\phi-\left[\left(\frac{\overline{\psi}\boldsymbol{\gamma}_{\rho}\psi
\overline{\psi}\boldsymbol{\gamma}^{\rho}\psi p^{3}}{16k^{2}}\right)^{\frac{1}{3}}
-\frac{\lambda}{4}v^{2}\right]\phi+\frac{Y}{2}\overline{\psi}\psi\approx0
\end{eqnarray}
and the vacuum configuration is
\begin{eqnarray}
&\lambda v^{2}\equiv\left.\left(\frac{4p^{3}\overline{\psi}\boldsymbol{\gamma}_{\rho}\psi
\overline{\psi}\boldsymbol{\gamma}^{\rho}\psi}{k^{2}}\right)^{\frac{1}{3}}\right|_{\mathrm{v}}
\end{eqnarray}
as it is straightforward to check; in the complementary approximation we get
\begin{eqnarray}
&i\boldsymbol{\gamma}^{\mu}\boldsymbol{\nabla}_{\mu}\psi
-\frac{9}{32pv^{2}}\overline{\psi}\boldsymbol{\gamma}_{\rho}\psi 
\boldsymbol{\gamma}^{\rho}\psi-Yv\psi\approx0
\end{eqnarray}
with scalar field equation given by
\begin{eqnarray}
&\nabla^{2}\phi-\left[\frac{9\overline{\psi}\boldsymbol{\gamma}_{\rho}\psi
\overline{\psi}\boldsymbol{\gamma}^{\rho}\psi}{64pv^{4}}
-\frac{\lambda}{4}v^{2}\right]\phi+\frac{Y}{2}\overline{\psi}\psi\approx0
\end{eqnarray}
and vacuum configuration as
\begin{eqnarray}
&\lambda v^{6}\equiv\left.\frac{9\overline{\psi}\boldsymbol{\gamma}_{\rho}\psi
\overline{\psi}\boldsymbol{\gamma}^{\rho}\psi}{16p}\right|_{\mathrm{v}}
\end{eqnarray}
once again as it is straightforward to check directly. Since a symmetry breaking mechanism has occurred we will now set $Yv\!=\!m$ in what follows.

We notice that in the high-density limit the fermionic field equations 
\begin{eqnarray}
&i\boldsymbol{\gamma}^{\mu}\boldsymbol{\nabla}_{\mu}\psi
-\left(\frac{27}{256k\overline{\psi}\boldsymbol{\gamma}_{i}\psi \overline{\psi}\boldsymbol{\gamma}^{i}\psi}\right)^{\frac{1}{3}}
\overline{\psi}\boldsymbol{\gamma}_{\rho}\psi \boldsymbol{\gamma}^{\rho}\psi
-m\psi\approx0
\label{high}
\end{eqnarray}
have torsionally-induced interactions scaling as the dynamic term and so they are renormalizable, while in the low-density limit fermionic field equations
\begin{eqnarray}
&i\boldsymbol{\gamma}^{\mu}\boldsymbol{\nabla}_{\mu}\psi
-\frac{9}{32pv^{2}}\overline{\psi}\boldsymbol{\gamma}_{\rho}\psi \boldsymbol{\gamma}^{\rho}\psi
-m\psi\approx0
\label{low}
\end{eqnarray}
have torsionally-induced spinorial interactions with the form they have in the Sciama-Kibble--Einstein gravity: this shows that even if the spin-contact fermion forces here are renormalizable they nevertheless approximate the spin-contact fermion forces we have in SKE gravitation, and consequently the non-linear Nambu-Jona--Lasinio potentials can be recovered; notice also that because the constants $k$ and $p$ must have the same sign then these spin-contact forces have the same features in both high-energy and low-energy limits, that is they are either always repulsive or always attractive. What this implies is that it is in fact possible to have a theory like the Nambu-Jona--Lasinio model working well as a low-energy approximation of a theory that in high-energy regimes becomes the renormalizable theory we have presented all along this paper.

As a further comment, we point out that in the limit in which the fermionic matter field tends to vanish then torsion vanishes too, showing that the torsion does not propagate, as we desired; this is crucial since the presence of torsion in vacuum has very stringent limits \cite{k-r-t}, and it is the reason for which we have asked no derivatives of torsion in the dynamics. We also have to notice that as it has been discussed in \cite{Fabbri}, general conformal models in vacuum do not necessarily have a torsionless limit and even when they do the limit of the field equations gives field equations that are restricted with respect to the purely metric gravitational field equations, but nevertheless as we have shown here in this conformal model the vacuum case is always torsionless and its field equations are those we would have had in the purely metric case precisely.

We know that such a process of mass generation already considered for fermions also generates the mass of the scalar as $M^{2}\!=\!\frac{1}{2}\lambda v^{2}$ as well as a cosmological constant given by $\Lambda\!=\!\frac{1}{16}\lambda v^{4}$ which in the high-energy limit are
\begin{eqnarray}
&M^{2}\equiv\left.\left(\frac{p^{3}\overline{\psi}\boldsymbol{\gamma}_{\rho}\psi
\overline{\psi}\boldsymbol{\gamma}^{\rho}\psi}{2k^{2}}\right)^{\frac{1}{3}}\right|_{\mathrm{v}}
\ \ \ \ \ \ \mathrm{and} \ \ \ \ \ \ 
\Lambda\equiv\left.\left(\frac{p^{3}\overline{\psi}\boldsymbol{\gamma}_{\rho}\psi
\overline{\psi}\boldsymbol{\gamma}^{\rho}\psi}{16k^{2}\sqrt{\lambda}^{3}}\right)^{\frac{2}{3}}\right|_{\mathrm{v}}
\label{H}
\end{eqnarray}
while in the low-energy limit they are
\begin{eqnarray}
&M^{2}\equiv\left.\left(\frac{9\lambda^{2}\ \overline{\psi}\boldsymbol{\gamma}_{\rho}\psi
\overline{\psi}\boldsymbol{\gamma}^{\rho}\psi}{128p}\right)^{\frac{1}{3}}\right|_{\mathrm{v}}
\ \ \ \ \ \ \mathrm{and} \ \ \ \ \ \ 
\Lambda\equiv\left.\left(\frac{9\sqrt{\lambda}\ \overline{\psi}\boldsymbol{\gamma}_{\rho}\psi
\overline{\psi}\boldsymbol{\gamma}^{\rho}\psi}{1024p}\right)^{\frac{2}{3}}\right|_{\mathrm{v}}
\label{L}
\end{eqnarray}
which are written in terms of the fermionic density of the vacuum alone and giving intriguing relationships: first of all we have to notice that $k$ and $p$ as well as $\lambda$ must be positive for consistency; however what we believe to be the most intriguing consequence of this construction is that the vacuum depends on the fermionic density and ultimately on the energy scale of the system.

For the moment we will carry out a qualitative analysis taking all coupling constants of the order of unity: at the scales of the standard model of particle physics one must employ (\ref{H}) giving the scalar mass $M\approx10^{2} \mathrm{GeV}$ as expected while there is no need to evaluate the cosmological constant because we cannot observed it by performing scattering of particles; on the other hand at the scale of the standard model of cosmology one must use (\ref{L}) for which there is no need to estimate the scalar mass since it cannot be detected in astrophysical experiments while the cosmological constant is $\Lambda\approx10^{-160}\mathrm{GeV}^{4}$ which is still not the observed value although in this model such a value is not \emph{much larger} but instead \emph{much smaller} than the observed one. This may solve the cosmological constant problem because, if on the one hand a large cosmological constant would need a still unknown additional mechanism followed by a further fine-tuning to quench its value, on the other hand a small cosmological constant does not need anything like this. All this happens in a qualitative analysis in which constants are normalized to unity, but this analysis may become quantitative if we allow the constants to have specific values: to this extent it is enough to notice that the high-energy form (\ref{H}) of the scalar mass depends on $p^{3}/k^{2}$ while the low-energy form (\ref{L}) of the cosmological constant depends on $\lambda/p^{2}$ showing not only that they are independent since $k$ and $\lambda$ are, but also that large values of $p$ would increase the value of the scalar mass diminishing the cosmological constant since $p$ appears in the numerator of the former and the denominator of the latter, as it is clear. And incidentally, this may also solve the coincidence problem stated as the correspondence between the cosmological constant and the density of the matter fields, which in the standard model of cosmology appears to be accidental; in the present approach they are in fact related by the cosmological constant formula (\ref{L}). This may change the treatment of these two problems of cosmology, showing how they might be solved at once.

To complete our work it is essential to recall that these problems related to Dark Energy have been faced here in terms of conformal gravity, but in conformal gravity Dark Energy topics might well receive a different treatment altogether; in fact, in a conformal theory the projective structure makes it entirely different from the standard one and a wholly new discussion must be carried out \cite{m1,m2,Mannheim:2005bfa}. It is then possible to think that both paths, that is the present model of torsional conformal cosmology and a different way to account for the projective structure of the observed universe, may be considered simultaneously, with new avenues that may be followed.
\section*{Conclusion}
In this paper, we considered axial torsion with no conformal transformations and we have insisted on the fact that no derivatives of the axial torsion should be present in the action while the metric was taken with the standard conformal transformation and implemented in the action in terms of the metric conformal curvature tensor, and gauge conformal fields were added; then massless spinors were also taken into account. The possibility to introduce massless scalar fields was considered and such fields were included. With this field content, a unique action invariant under conformal transformations was obtained, yielding field equations in which torsion appeared algebraically, so that the axial torsion coupled to the axial spin density of the spinor field and torsion could be substituted in terms of the spin density of the Dirac field: the Weyl field equations reduced to the usual form plus non-linear contributions while the Maxwell field equations were essentially unchanged, and the Dirac matter field equations contained non-linear terms as well; the scalar fields received non-linear contributions as desired for the symmetry breaking mechanism. We have found that: the Dirac matter field equations had non-linear terms that in the ultraviolet limit were renormalizable but which in the infrared limit reduced to the Nambu-Jona--Lasinio potentials; the case of stable stationary breaking of the symmetry was given by non-vanishing vacua that were not universal constants, nor several different parameters, but the density of fermionic fields, so that the mechanism of scalar mass and cosmological constant generation works as usual, but now different energy scales can have different vacua resulting into different values for the scalar mass and the cosmological constant. This introduces a new way to meet the cosmological constant and coincidence problems within a single framework.

Imagining the universe to have fermionic particles wiggling about in a bath of a scalar field filling the universe itself, we can figure fermions interacting with the scalar and receiving their masses as usual but now they also affect back the vacuum of the scalar and so its mass: here not only the fermions receive mass because of their interaction with the scalar but also the scalar receives mass because of its interaction with the fermions: in particle physics or cosmology of the early universe, the mass generation mechanism works as usual, although now we have a dynamical symmetry breaking; in cosmological models that involve the recent universe, the fermion field distribution decreases and correspondingly the mass of the scalar particle excitations decreases, but since in cosmological experiments the mass of the scalar is manifested as cosmological constant then the cosmological constant must be small too. How small this value is depends on the precise value of the scalar mass in cosmology and then on the low-density distribution of fermions in the universe; universes voided of fermions have massless scalars and no cosmological constant. This is an unexplored way to solve both cosmological constant and coincidence problem simultaneously.

Combining these ideas with pre-existing ones by Mannheim may allow us to think anew the way we might approach these cosmological issues.

\

\noindent\textbf{Acknowledgments.} I would like to thank Professor Philip D.~Mannheim for the interesting discussions we had about the role of the projective structure of conformal gravity and its possible applications.


\begin{thebibliography}{40}
\bibitem{m-t-w}
C.~Misner, K.~Thorne, J.~A.~Wheeler, 
\textit{Gravitation} (Freeman, 1973).
\bibitem{a-l}
J.~Audretsch, C.~L\"{a}mmerzahl,
\textit{Class. Quant. Grav.} \textbf{5}, 1285 (1988).
\bibitem{m-l}
A.~Macias, C.~L\"{a}mmerzahl,
\textit{J. Math. Phys.} \textbf{34}, 4540 (1993).
\bibitem{xy}
Xin Yu, 
\textit{Astrophysical and Space Science} \textbf{154}, 321 (1989).
\bibitem{so}
M.~Socolovsky,
\textit{Annales Fond. Broglie} \textbf{37}, 73 (2012).
\bibitem{F1}
L.~Fabbri,
in \textit{Annales de la Fondation de Broglie: Special 
Issue on Torsion} (Ed. Dvoeglazov, Fondation de Broglie, 2007).
\bibitem{sh}
I.~L.~Shapiro,
\textit{Phys. Rept.} \textbf{357}, 113 (2002).
\bibitem{fabbri}
L.~Fabbri
\textit{Phys. Lett. B} \textbf{707}, 415 (2012).
\bibitem{e-p-s}
J.~Ehlers, F.~Pirani, A.~Schild, in \textit{Synge Festschrift}\\
(Oxford University Press, 1972).
\bibitem{h-h-k-n}
F.~W.~Hehl, P.~Von Der Heyde, G.~D.~Kerlick, J.~M.~Nester,\\
\textit{Rev. Mod. Phys.} \textbf{48}, 393 (1976).
\bibitem{F2}
L.~Fabbri,
\textit{Annales Fond. Broglie} \textbf{33}, 365 (2008).
\bibitem{s}
K.~S.~Stelle,
\textit{Phys. Rev. D} \textbf{16}, 953 (1977).
\bibitem{Fabbri:2011ha} 
L.~Fabbri,
arXiv:1101.2334 [gr-qc].
\bibitem{Fatibene:1998aa} 
L.~Fatibene, M.~Francaviglia,
\textit{Acta Phys. Polon. B} \textbf{29}, 915 (1998).
\bibitem{Fatibene:1996mc} 
L.~Fatibene, M.~Ferraris, M.~Francaviglia, M.~Godina,\\
\textit{Gen. Rel. Grav.} \textbf{30}, 1371 (1998).
\bibitem{k-r-t}
V.~A.~Kostelecky, N.~Russell, J.~Tasson,\\
\textit{Phys. Rev. Lett.} \textbf{100}, 111102 (2008).
\bibitem{Fabbri}
L.~Fabbri,
arXiv:1104.5002 [gr-qc].
\bibitem{m1}
P.~D.~Mannheim,
\textit{Gen. Rel. Grav.} \textbf{22}, 289 (1990).
\bibitem{m2}
P.~D.~Mannheim, 
\textit{Gen. Rel. Grav.} \textbf{43}, 703 (2011).
\bibitem{Mannheim:2005bfa} 
P.~D.~Mannheim,
\textit{Prog. Part. Nucl. Phys.} \textbf{56}, 340 (2006).
\end{thebibliography}
\end{document}